\documentclass[conference]{IEEEtran}
\IEEEoverridecommandlockouts
\usepackage{cite}
\usepackage{amsmath,amssymb,amsfonts}
\usepackage{algorithmic}
\usepackage{graphicx}
\usepackage{textcomp}
\usepackage[table]{xcolor}
\usepackage{listings}
\usepackage{tabularx}
\newcolumntype{L}{>{\raggedright\arraybackslash}X}
\usepackage{fancyvrb}
\usepackage{lmodern}
\usepackage{ifxetex,ifluatex}
\def\BibTeX{{\rm B\kern-.05em{\sc i\kern-.025em b}\kern-.08em
    T\kern-.1667em\lower.7ex\hbox{E}\kern-.125emX}}

\usepackage{lmodern}
\usepackage{amssymb,amsmath}
\usepackage{ifxetex,ifluatex}
\ifnum 0\ifxetex 1\fi\ifluatex 1\fi=0 % if pdftex
  \usepackage[T1]{fontenc}
  \usepackage[utf8]{inputenc}
  \usepackage{textcomp} % provide euro and other symbols
\else % if luatex or xetex
  \usepackage{unicode-math}
  \defaultfontfeatures{Scale=MatchLowercase}
  \defaultfontfeatures[\rmfamily]{Ligatures=TeX,Scale=1}
\fi
\IfFileExists{upquote.sty}{\usepackage{upquote}}{}
\IfFileExists{microtype.sty}{% use microtype if available
  \usepackage[]{microtype}
  \UseMicrotypeSet[protrusion]{basicmath} % disable protrusion for tt fonts
}{}
\makeatletter
\@ifundefined{KOMAClassName}{% if non-KOMA class
  \IfFileExists{parskip.sty}{%
    \usepackage{parskip}
  }{% else
    \setlength{\parindent}{0pt}
    \setlength{\parskip}{6pt plus 2pt minus 1pt}}
}{% if KOMA class
  \KOMAoptions{parskip=half}}
\makeatother
\usepackage{xcolor}
\IfFileExists{xurl.sty}{\usepackage{xurl}}{} % add URL line breaks if available
\IfFileExists{bookmark.sty}{\usepackage{bookmark}}{\usepackage{hyperref}}
\hypersetup{
  hidelinks,
  pdfcreator={LaTeX via pandoc}}
\usepackage{color}
\usepackage{fancyvrb}

\DefineVerbatimEnvironment{Highlighting}{Verbatim}{commandchars=\\\{\},fontsize=\small}
\newenvironment{Shaded}{}{}

\newcommand{\AttributeTok}[1]{\textcolor[rgb]{0.49,0.56,0.16}{#1}}

\newcommand{\CommentTok}[1]{\textcolor[rgb]{0.38,0.63,0.69}{\textit{#1}}}

\newcommand{\ControlFlowTok}[1]{\textcolor[rgb]{0.00,0.44,0.13}{\textbf{#1}}}
\newcommand{\DataTypeTok}[1]{\textcolor[rgb]{0.56,0.13,0.00}{#1}}
\newcommand{\DecValTok}[1]{\textcolor[rgb]{0.25,0.63,0.44}{#1}}

\newcommand{\ExtensionTok}[1]{#1}
\newcommand{\FloatTok}[1]{\textcolor[rgb]{0.25,0.63,0.44}{#1}}

\newcommand{\KeywordTok}[1]{\textcolor[rgb]{0.00,0.44,0.13}{\textbf{#1}}}
\newcommand{\NormalTok}[1]{#1}

\newcommand{\SpecialCharTok}[1]{\textcolor[rgb]{0.25,0.44,0.63}{#1}}

\newcommand{\StringTok}[1]{\textcolor[rgb]{0.25,0.44,0.63}{#1}}

\providecommand{\tightlist}{%
  \setlength{\itemsep}{0pt}\setlength{\parskip}{0pt}}
\begin{document}

\title{
mdspan in C++: A Case Study in the Integration of Performance Portable Features into International Language Standards
\thanks{This work was carried out in part at Sandia National Laboratories. Sandia National Laboratories is a multi-mission laboratory managed and operated by National Technology \& Engineering Solutions of Sandia, LLC, a wholly owned subsidary of Honeywell International Inc., for the U. S. Department of Energy’s National Nuclear Security Administration under contract DE-NA0003525.  This document is approved for unclassified unlimited release as SAND2019-10501 C. }
}

\author{\IEEEauthorblockN{\ \ \ \ D. S. Hollman\ \ \ \ }
\IEEEauthorblockA{\textit{Scalable Modeling \& Analysis} \\
\textit{Sandia National Laboratories}\\
Livermore, CA, USA \\
dshollm@sandia.gov}
\and
\IEEEauthorblockN{Bryce Adelstein Lelbach}
\IEEEauthorblockA{\ \ \ \ \ \ \ \ \ \ \ \textit{NVIDIA Corp.}\ \ \ \ \ \ \ \ \ \ \  \\
Santa Clara, CA, USA \\
brycelelbach@gmail.com}
\and
\IEEEauthorblockN{\ \ \ \ \ \ \ \ \ \ \ \ H. Carter Edwards\ \ \ \ \ \ \ \ \ \ \ \ }
\IEEEauthorblockA{\textit{NVIDIA Corp.} \\
Santa Clara, CA, USA \\
hedwards@nvidia.com}
\and
\IEEEauthorblockN{Mark Hoemmen}
\IEEEauthorblockA{\textit{Engineering Sciences} \\
\ \ \ \ \textit{Sandia National Laboratories}\ \ \ \ \\
Albuquerque, NM, USA \\
mhoemme@sandia.gov}
\and
\IEEEauthorblockN{Daniel Sunderland}
\IEEEauthorblockA{\textit{Center for Computing Research} \\
\textit{Sandia National Laboratories}\\
Albuquerque, NM, USA \\
dsunder@sandia.gov}
\and
\IEEEauthorblockN{Christian R. Trott}
\IEEEauthorblockA{\textit{Center for Computing Research} \\
\textit{Sandia National Laboratories}\\
Albuquerque, NM, USA \\
crtrott@sandia.gov}
}

\maketitle

\begin{abstract}
Multi-dimensional arrays are ubiquitous in high-performance computing
(HPC), but their absence from the C++ language standard is a
long-standing and well-known limitation of their use for HPC. This paper
describes the design and implementation of \texttt{mdspan}, a proposed
C++ standard multidimensional array view (planned for inclusion in
C++23). The proposal is largely inspired by work done in the Kokkos
project---a C++ performance-portable programming model deployed by
numerous HPC institutions to prepare their code base for exascale-class
supercomputing systems. This paper describes the final design of mdspan
after a five-year process to achieve consensus in the C++ community. In
particular, we will lay out how the design addresses some of the core
challenges of performance-portable programming, and how its
customization points allow a seamless extension into areas not currently
addressed by the C++ Standard but which are of critical importance in
the heterogeneous computing world of today's systems. Finally, we have
provided a production-quality implementation of the proposal in its
current form. This work includes several benchmarks of this
implementation aimed at demonstrating the zero-overhead nature of the
modern design.

\end{abstract}

\begin{IEEEkeywords}
programming models, C++, performance portability, programming languages, stardardization, multidimensional array, data structures
\end{IEEEkeywords}

%================================================================================%
\section{Introduction}

One of the primary concerns of the high-performance computing (HPC)
community\cite{DOEPPP} is performance portability. \emph{Performance
portability} means that a single code base can perform well on many
different platforms. Over the last decade in particular, numerous
projects\cite{gridcpp2018,alpaka2016,occa2014,raja2014,kokkos2014} have
tried to address various challenges associated with it. The recent
announcement of the first exascale-class platforms, introducing
architectures which were previously not deployed in the HPC community,
has increased the urgency of finding solutions to performance
portability concerns. One of the projects which has found significant
success in adoption is Kokkos,\cite{kokkos2014,kokkosgithub} a C++
performance-portable programming model originally developed at Sandia
National Laboratories, but now maintained by a group spanning four
United States National Laboratories as well as the Swiss National
Supercomputing Centre.

Arguably the most significant innovation of the Kokkos project was its
\texttt{View} data structure, a multi-dimensional array abstraction
which addresses concerns of performance portability such as data layout
and data access customization. This array abstraction is now used at the
heart of many HPC software projects\cite{kokkosprojects}, and is proving
to be critical for meeting the challenges of preparing code bases for
the exascale era. While maintaining these capabilities in an
HPC-specific solution is workable for now, there are a number of reasons
why it would be beneficial to have the core capabilities become part of
international programming language standards. Doing so would enable
tighter integration into other language and library capabilities, such
as the proposed ISO C++ Linear Algebra library\cite{wg21_p1673}. It
would simplify interface compatibility between different HPC products,
and would further seamless integration with external products used in
applications not specific to HPC. For example, the proposed ISO C++
Audio library\cite{wg21_p1386} has expressed interest in using this
abstraction.

To that end the Kokkos team initiated a collaboration with other
stakeholders to design a multi-dimensional array for the ISO C++
Standard, that also addresses the concerns of performance portability
addressed by the Kokkos \texttt{View} abstraction. The result of this
over five-year process is \texttt{std::mdspan}, described herein and
proposed to the ISO C++ standard in the proposal P0009\cite{wg21_p0009}.
The design allows for a mix of static and dynamic array dimensions,
enables control of the data layout, and has customization points to
control how data are accessed. The latter includes use cases that
involve hardware-specific special load and store paths.

In this work we describe each of the design aspects of \texttt{mdspan},
with examples demonstrating their impact for performance and portability
concerns, as well as benchmarks of the production-quality reference
implementation developed by the authors.

%================================================================================%

%================================================================================%
\section{Design}

\texttt{mdspan} provides a class template for creating types of objects
that represent, but do not own, a contiguous piece (or ``span'') of
memory that is to be treated as a multi-dimensional entity with one or
more dimensional constraints. Together, these dimensional constraints
form a \emph{multi-index domain}. In the simple case of a
two-dimensional entity, for instance, this multi-index domain
encompasses the row and column indices of what is typically called a
matrix. For instance,

\begin{Shaded}
\begin{Highlighting}[]
\DataTypeTok{void}\NormalTok{ some\_function(}\DataTypeTok{float}\NormalTok{* data) \{}
  \KeywordTok{auto}\NormalTok{ my\_matrix =}
\NormalTok{    mdspan\textless{}}\DataTypeTok{float}\NormalTok{, dynamic\_extent, dynamic\_extent\textgreater{}(}
\NormalTok{      data, }\DecValTok{20}\NormalTok{, }\DecValTok{40}
\NormalTok{    );}
  \CommentTok{/* ... */}
\NormalTok{\}}
\end{Highlighting}
\end{Shaded}

says to create an object that interprets memory starting at the pointer
\texttt{data} as a matrix with the shape 20 rows by 40 columns. Extents
can be provided either statically (i.e., at compile time) or dynamically
(as shown above), and static extents can be mixed with dynamic extents:

\begin{Shaded}
\begin{Highlighting}[]
\DataTypeTok{void}\NormalTok{ another\_function(}\DataTypeTok{float}\NormalTok{* data) \{}
  \KeywordTok{auto}\NormalTok{ another\_matrix =}
\NormalTok{    mdspan\textless{}}\DataTypeTok{float}\NormalTok{, }\DecValTok{20}\NormalTok{, dynamic\_extent\textgreater{}(}
\NormalTok{      data, }\DecValTok{40}
\NormalTok{    );}
  \CommentTok{/* ... */}
\NormalTok{\}}
\end{Highlighting}
\end{Shaded}

This code snippet also treats \texttt{data} as a 20 by 40 matrix, but
the first of these dimensions is ``baked in'' to the type at compile
time---all instances of the type
\texttt{mdspan\textless{}double,\ 20,\ dynamic\_extent\textgreater{}}
will have 20 rows.

The design is greatly simplified by delegating the ownership and
lifetime management of the data to orthogonal constructs. Thus,
\texttt{mdspan} merely interprets existing memory as a multi-dimensional
entity, leaving management of the underlying memory to the user. This
follows a trend of similar constructs recently introduced to C++, such
as
\texttt{string\_view}\cite{wg21_is_14882:2017, cppreference_string_view}
and \texttt{span}\cite{wg21_is_14882:2020_cd, cppreference_span}. These
constructs allow ``API funnelling,'' which makes it easy for libraries
to support users' own types instead of forcing users to use a specific
type. Library interfaces can take \texttt{string\_view}s or
\texttt{mdspan}s, and library users can add interfaces to their own
types that return a suitable \texttt{string\_view} or \texttt{mdspan}.
This design pattern enables easy adoption by existing codebases which
have their own matrix types. Since \texttt{mdspan} is non-owning, users
can always create an \texttt{mdspan} that refers to a matrix owned by
another object. Older abstractions also take this approach. Iterators,
which have been central to C++ algorithm design for decades, are also
non-owning entities which delegate lifetime management as a separate
concern.\cite{stepanov2009}

References to entries in these matrices are obtained by giving a
multi-index (that is, an ordered set of indices) to the object's
\texttt{operator()}, which has been overloaded for this purpose:

\begin{Shaded}
\begin{Highlighting}[]
\CommentTok{// add 3.14 to the value on the row with index 10}
\CommentTok{// and the column with index 5}
\NormalTok{some\_matrix(}\DecValTok{10}\NormalTok{, }\DecValTok{5}\NormalTok{) += }\FloatTok{3.14}\NormalTok{;}
\CommentTok{// print the value of the entry in the row with}
\CommentTok{// index 0 and the column with index 38}
\NormalTok{printf(}\StringTok{"}\SpecialCharTok{\%f}\StringTok{"}\NormalTok{, some\_matrix(}\DecValTok{0}\NormalTok{, }\DecValTok{38}\NormalTok{));}
\end{Highlighting}
\end{Shaded}

The length of each dimension is accessed via the \texttt{extent} member
function. It takes an index to indicate the dimension. A loop to
multiply all entries of the matrix by a scalar could thus look like
this:

\begin{Shaded}
\begin{Highlighting}[]
\ControlFlowTok{for}\NormalTok{(}\DataTypeTok{int}\NormalTok{ row = }\DecValTok{0}\NormalTok{; row \textless{} my\_mat.extent(}\DecValTok{0}\NormalTok{); ++row)}
  \ControlFlowTok{for}\NormalTok{(}\DataTypeTok{int}\NormalTok{ col = }\DecValTok{0}\NormalTok{; col \textless{} my\_mat.extent(}\DecValTok{1}\NormalTok{); ++col)}
\NormalTok{    my\_mat(row, col) *= }\FloatTok{2.0}\NormalTok{;}
\end{Highlighting}
\end{Shaded}

Arbitrary slices of an \texttt{mdspan} can be taken using the
\texttt{subspan} function:

\begin{Shaded}
\begin{Highlighting}[]
\KeywordTok{auto}\NormalTok{ my\_tens = mdspan\textless{}}\DataTypeTok{float}\NormalTok{, }\DecValTok{3}\NormalTok{, }\DecValTok{4}\NormalTok{, }\DecValTok{5}\NormalTok{, }\DecValTok{20}\NormalTok{\textgreater{}(data);}
\KeywordTok{auto}\NormalTok{ my\_matrix = subspan(my\_tens,}
  \DecValTok{2}\NormalTok{, all, pair\{}\DecValTok{2}\NormalTok{, }\DecValTok{4}\NormalTok{\}, }\DecValTok{0}
\NormalTok{);}
\end{Highlighting}
\end{Shaded}

The above snippet creates a 4 by 2 matrix sub-view of \texttt{my\_tens}
where the entries \texttt{i,\ j} correspond to index 2 in the first
dimension of \texttt{my\_tens}, index \texttt{i} in the second
dimension, \texttt{j+2} in the third dimension, and \texttt{0} in the
fourth dimension. This relatively verbose syntax for slicing was
preferred over other approaches, because slicing needs can vary
substantially across different domains. Domain-specific syntax can
easily be built on top of \texttt{subspan}.

Just as \texttt{std::string} is actually a C++ alias for
\texttt{std::basic\_string}\cite{wg21_is_14882:2017,cppreference_string_view},
\texttt{std::mdspan} is an alias for \texttt{std::basic\_mdspan}.
Whereas \texttt{std::mdspan} only provides control over the scalar type
and the extents, \texttt{std::basic\_mdspan} exposes more customization
points. It is templated on four parameters: the scalar type, the extents
object, the layout, and the accessor. In the following sections, we will
describe these parameters and their value for improving performance or
increasing portability.

\subsection{Extents Class Template}

In \texttt{basic\_mdspan} the extents are provided via an
\texttt{extents} class template. As with the \texttt{mdspan} alias
template, the parameters are either static sizes or the
\texttt{dynamic\_extent} tag.

\begin{Shaded}
\begin{Highlighting}[]
\DataTypeTok{void}\NormalTok{ some\_function(}\DataTypeTok{float}\NormalTok{* data) \{}
  \KeywordTok{auto}\NormalTok{ my\_matrix =}
\NormalTok{    basic\_mdspan\textless{}}\DataTypeTok{float}\NormalTok{, extents\textless{}}\DecValTok{20}\NormalTok{, dynamic\_extent\textgreater{}\textgreater{}(}
\NormalTok{      data, }\DecValTok{40}
\NormalTok{    );}
  \CommentTok{/* ... */}
\NormalTok{\}}
\end{Highlighting}
\end{Shaded}

The ability to provide extents statically can help significantly with
compiler optimizations. For example, a compiler may be able to unroll
small inner loops completely if the extents are known at compile time.
Knowing exact counts and sizes can also help with vectorization and the
optimizer's cost model. A typical example of this in HPC is operations
on a batch of small matrices or vectors, where the dimensions of each
item are dictated by a physics property or the way the system was
discretized, rather than by the problem size. When this sort of problem
interacts with generic code, such information would be lost unless
static extents can be part of the \texttt{mdspan} type itself. The
\texttt{TinyMatrixSum} benchmark (below) provides a proxy for problems
with this sort of behavior.

\subsection{Layout abstraction}

Modern C++ design requires library authors to orthogonalize certain
aspects of the design into customization points, over which algorithms
may be written generically. The most commonplace example of this is the
\texttt{Allocator}
abstraction\cite{wg21_is_14882:2017,cppreference_allocator_ntr}, which
controls memory allocation for standard containers like
\texttt{std::vector}\cite{wg21_is_14882:2017,cppreference_vector}. Most
algorithms on containers do not change regardless of how the underlying
data is allocated. The \texttt{Allocator} abstraction allows such
algorithms to be generic over the form of memory allocation used by the
container.

An example of one such aspect in the current context is the layout of
the underlying data with respect to the multi-index domain. While a
high-quality-of-implementation matrix multiply would definitely
specialize for different data layouts, the simplest possible
implementation would only need to know how to get and store data
associated with a given multi-index into the underlying memory. This
also describes the majority of use cases from the perspective of the
caller of such algorithms, where only the semantics of a mathematical
matrix multiply are needed regardless of data layout. The grouping of a
single set of mathematical semantics under a common algorithm name
(regardless of layout) serves as a conduit for performance portability,
and additionally reduces the cognitive load for the writer and
particularly the reader of the code.

The canonical example, again with reference to data layout, is the
portability of access patterns in code that may run on a
latency-optimizer processed (e.g., CPU) or on a bandwidth-optimized
processor (e.g., GPU). GPUs need to coalesce accesses (that is, stride
across execution agents) because of the vector nature of the underlying
hardware, whereas CPUs want to maximize locality (that is, assign
contiguous chunks to the same execution agent) in order to increase
cache reuse.

The abstraction for representing data layout generically is called the
\texttt{LayoutMapping}. The primary task of the \texttt{LayoutMapping}
is to represent the transformation of a multi-index into a single,
scalar memory offset. A large number of algorithms on multi-dimensional
arrays have semantics that depend only on the data as retrieved through
the multi-index domain, indicating that this transformation is a prime
aspect for orthogonalization into a customization point. (Note that many
algorithms have \emph{performance} characteristics that depend on this
transformation, but the separation of semantic aspects of an algorithm
from its performance characteristics is critical to modern programming
model design. The fact that the \texttt{LayoutMapping} abstraction
promotes this separation is further evidence of its utility as a
customization point.)

A brief survey of existing practice (such as the BLAS Technical Standard
\cite{BLAS}, Eigen\cite{eigenweb}, and MAGMA\cite{magma}) reveals an
initial set of layout mappings that such an abstraction must support.
These include, at minimum,

\begin{itemize}
\tightlist
\item
  row-major or column-major layouts (represented by the \texttt{TRANS}
  parameters in BLAS), that generalize to describe layouts where the
  fast-running index is left-most or right-most;
\item
  strided layouts (represented by the \texttt{LD} parameters in BLAS),
  that generalize to any in a class of layouts that can describe the
  distance in memory between two consecutive indices in a particular
  dimension with a constant (specific to that dimension); and
\item
  symmetric layouts (e.g., from the \texttt{xSYMM} algorithms in BLAS),
  which also include generalizations like whether the upper or lower
  triangle is stored (the \texttt{UPLO} parameter in BLAS) and whether
  the diagonal is stored explicitly, implicitly, or in some separate,
  contiguous storage.
\end{itemize}

In addition to similarities, it also helps to look at what differences
these layout mappings may introduce, over which some algorithms may not
be generic. In general, as many previous researchers have
noted,\cite{sutton2011design} the design of generic concepts for
customization typically begins with the algorithms, not the data
structures. Much of the design of \texttt{LayoutMapping} can be
motivated with some very simple algorithms. Consider an algorithm,
\texttt{scale}, that takes an \texttt{mdspan} and a scalar and
multiplies each entry, in place, by the scalar. For brevity, we will
only consider the two-dimensional case here (though much of this
motivation can be done even in the one-dimensional case). If such an
algorithm is to be implemented in the simplest possible way---iterating
over the rows and column indices and scaling each element---the
implementation would fail to meet the semantic requirements of the
algorithm for symmetric layouts, since non-diagonal entries reference
the same memory. Thus, it is necessary for certain algorithms to know
whether each multi-index in the domain maps to a unique offset in the
codomain (the space of all offsets that could be valid results of the
mapping). (An example of an algorithm for which this requirement is
\emph{not} needed is \texttt{dot\_product}.) The \texttt{LayoutMapping}
customization expresses this property through the requirement that it
provide an \texttt{is\_unique} method. Many algorithms are difficult or
impossible to implement on general non-unique layouts. However, in the
simple case of \texttt{scale}, the algorithm \emph{could} be implemented
for any layout that is simply \emph{contiguous}, by viewing the codomain
of the layout as a one-dimensional \texttt{mdspan} and scaling each item
that way. Contiguousness is expressed through the requirement of an
\texttt{is\_contiguous} method, and the size of the codomain is
expressed through the \texttt{required\_span\_size} required method.
Similarly, as previously observed, many existing implementations (such
as the BLAS) can specially handle any layout with regular strides.
Layout mappings can express whether they are strided using the
\texttt{is\_strided} method. Finally, all of these aspects need to be
expressible statically and dynamically, so for layout mappings where the
uniqueness, stridedness, and continguousness are consistently
\texttt{true} for all instances of the type, the
\texttt{is\_always\_unique}, \texttt{is\_always\_strided}, and
\texttt{is\_always\_contiguous} hooks are provided in the concept. These
requirements allow, for instance, algorithms that cannot support layouts
lacking certain properties to fail at compile time rather than run time.
The requirements on the \texttt{LayoutMapping} concept are summarized in
Table \ref{layoutreqs}.

\begin{table}[htbp]
\caption{Requirements on the \texttt{LayoutMapping} Concept}
\begin{center}
\begin{tabular}{|c|p{15em}|}
\hline
Expression & Meaning \\
\hline
    \texttt{M} & A \texttt{LayoutMapping} type. \\
    \texttt{m} & An instance of \texttt{M}. \\
    \texttt{E} & A specialization of \texttt{std::extents}. \\
    \texttt{e} & An instance of \texttt{E}. \\
    \texttt{i...} and \texttt{j...} & Multidimensional indices in the multidimensional index space described by \texttt{e}. \\
    \texttt{m.extents()} & The extents object \texttt{e} representing the multidimensional index domain of the mapping.  \\
    \texttt{m(i...)} & A nonnegative value representing the codomain offset corresponding to \texttt{i...}. \\
    \texttt{m.required\_span\_size()} & The maximum value of \texttt{m(i...)} plus \texttt{1} if all extents are non-zero, or \texttt{0} otherwise. \\
    \texttt{m.is\_unique()} & \texttt{true} only if for every \texttt{i... != j...}, \texttt{m(i...) != m(j...)}. \\
    \texttt{m.is\_contiguous()} & \texttt{true} only if the set defined by all \texttt{m(i...)} equals the set $\{0, ..., \mathtt{m.required\_span\_size()} - 1\}$. \\
    \texttt{m.is\_strided()} & \texttt{true} only if $\forall r \in [0, \mathtt{e.rank()}), \exists K_r$ such that $\forall \mathtt{i...}, \mathtt{j...} \in \mathtt{e}$, if all elements of \texttt{i...} and \texttt{j...} are equal except for the $r^{\mathrm{th}}$ element, with $j_r = i_r + 1$, then $K_r$ equals \texttt{m(j...) - m(i...)}. \\
    \texttt{m.stride(r)} & The integer $K_r$, as described above.  Only required if \texttt{m.is\_strided()} is \texttt{true}. \\
    \texttt{M::is\_always\_unique()} & \texttt{true} only if \texttt{m.is\_unique()} is \texttt{true} for all instances of \texttt{M}. \\
    \texttt{M::is\_always\_contiguous()} & \texttt{true} only if \texttt{m.is\_contiguous()} is \texttt{true} for all instances of \texttt{M}. \\
    \texttt{M::is\_always\_strided()} & \texttt{true} only if \texttt{m.is\_strided()} is \texttt{true} for all instances of \texttt{M}. \\
\hline
\end{tabular}

\label{layoutreqs}
\end{center}
\end{table}

\subsection{Accessor abstraction}

After several design iterations,\cite{wg21_p0009} the authors came to
the conclusion that many of the remaining customizations could be
encapsulated in the answer to one question: how should the
implementation turn an instance of some pointer type and an offset
(obtained from the \texttt{LayoutMapping} abstraction) into an instance
of some reference type? The \texttt{Accessor} customization point is
designed to provide all of the necessary flexibility in the answer to
this question. Our exploration in this space began with a couple of
specific use cases: a non-aliasing \texttt{Accessor}, similar to the
\texttt{restrict} keyword in C,\cite{c18standard} and an atomic
\texttt{Accessor}, where operations on the resulting reference use
atomic operations. The former needs to customize the pointer type to
include implementation-specific annotations (usually some variant of the
C-style \texttt{restrict} keyword) that indicate the pointer does not
alias pointers derived from other sources within the same context
(usually a function scope). The latter needs to customize the reference
type produced by the dereference operation to have it return a
\texttt{std::atomic\_ref\textless{}T\textgreater{}}.
(\texttt{std::atomic\_ref\textless{}T\textgreater{}} was merged into the
C++ Standard working draft during the C++20 cycle, and will likely be
officially approved as part of the C++20 balloting process when that
process completes sometime in 2020\cite{wg21_p0019}.) These requirements
immediately led us to include customizable \texttt{reference} and
\texttt{pointer} type names as part of the \texttt{Accessor} concept.
Marrying these two customizations could take several forms. One
possibility is to have a function that simply takes a \texttt{pointer}
and returns a \texttt{reference}. However, this requires the
\texttt{pointer} type to be arbitrarily offsettable---e.g., using
\texttt{operator+} or \texttt{std::advance}. A simpler approach that
removes this requirement is to have a customization point that takes the
\texttt{pointer} and an offset and returns the \texttt{reference}
directly. We chose the latter in order to simplify the requirements on
the \texttt{pointer} type, and named this required method
\texttt{access}.

The issue of offsetting a \texttt{pointer} to create another
\texttt{pointer}, while not necessarily separable from the creation of a
\texttt{reference}, is nonetheless also a concern that \texttt{Accessor}
needs to address for the implementation of the \texttt{subspan}
function. We named this customization with a required method
\texttt{offset}. The type of the \texttt{pointer} retrieved when
arbitrarily offsetting a \texttt{pointer} type may not necessarily match
the input pointer type. For instance, in the case of an overaligned
pointer type used for easy vectorization, a pointer derived from an
arbitrary (runtime) offset to this pointer cannot guarantee the
preservation of this alignment. Thus, the \texttt{Accessor} is allowed
to provide a different \texttt{Accessor}, named with the required type
name \texttt{offset\_policy}, that differs in type from itself (and
thus, for instance, may differ in its \texttt{pointer} type). Finally,
given an arbitrary \texttt{pointer} type, the current design requires
the ability to ``decay'' this type into an ``ordinary'' C++ pointer for
compatibility with \texttt{std::span}, which does not support
\texttt{pointer} type customization. The requirements on the
\texttt{Accessor} concept are summarized in Table \ref{accessreqs}.

\begin{table}[htbp]
\caption{Requirements on the \texttt{Accessor} Concept}
\begin{center}
\begin{tabular}{|c|p{15em}|}
\hline
Expression & Meaning \\
\hline
    \texttt{A} & An \texttt{Accessor} type. \\
    \texttt{a} & An instance of \texttt{A}. \\
    \texttt{A::element\_type} & The type of each element in the set of elements described by the associated \texttt{mdspan}. \\
    \texttt{A::pointer} & The pointer type through which a range of elements are accessed. \\
    \texttt{p} & An instance of \texttt{A::pointer}. \\
    \texttt{i} & Non-negative value of type \texttt{ptrdiff\_t}. \\
    \texttt{A::reference} & The type through which an element is accessed.  Must be convertible to \texttt{A::element\_type}. \\
    \texttt{A::offset\_policy} & (Optional) An \texttt{Accessor} type convertible from \texttt{A}. Defaults to \texttt{A}. \\
    \texttt{a.access(p, i)} & Returns an object that provides access to the \texttt{i}-th element in the range of elements that starts at \texttt{p}. \\
    \texttt{a.offset(p, i)} & An instance of \texttt{A::offset\_policy::pointer} for which \texttt{A::offset\_policy(a).access( a.offset(p, i), 0)} references the same element as \texttt{a.access(p, i)}. \\
\hline
\end{tabular}

\label{accessreqs}
\end{center}
\end{table}

\subsubsection{Accessor Use Case: Non-Aliasing Semantics}

As a concrete example, the (trivial) \texttt{Accessor} required to
express non-aliasing semantics (similar to the \texttt{restrict} keyword
and supported in many C++ compilers as \texttt{\_\_restrict}) is shown
in Figure \ref{restrict-accessor}. This differs from the default
accessor (\texttt{std::accessor\_basic\textless{}T\textgreater{}}) only
in the definition of the nested type \texttt{pointer}. Interestingly,
because the design of \texttt{mdspan} requires the \texttt{pointer} to
be used as a parameter (in \texttt{access}) before it is ever turned
into a reference, \texttt{mdspan} is able to skirt the well-known issues
surrounding the meaning of the \texttt{restrict} qualifier on a data
member of a
struct.\cite{gcc_docs_restrict,msvc_docs_restrict,finkel2014restrict}

\begin{figure}[!h]

\begin{Shaded}
\begin{Highlighting}[]
\KeywordTok{template}\NormalTok{ \textless{}}\KeywordTok{class}\NormalTok{ T\textgreater{}}
\KeywordTok{struct}\NormalTok{ RestrictAccessor \{}
  \KeywordTok{using} \DataTypeTok{element\_type}\NormalTok{ = T;}
  \KeywordTok{using}\NormalTok{ pointer = T* }\ExtensionTok{\_\_restrict}\NormalTok{;}
  \KeywordTok{using}\NormalTok{ reference = T\&;}
\NormalTok{  reference access(pointer p, }\DataTypeTok{ptrdiff\_t}\NormalTok{ i)}
    \AttributeTok{const} \KeywordTok{noexcept}
\NormalTok{  \{ }\ControlFlowTok{return}\NormalTok{ p[i]; \}}
\NormalTok{  pointer offset(pointer p, }\DataTypeTok{ptrdiff\_t}\NormalTok{ i)}
    \AttributeTok{const} \KeywordTok{noexcept}
\NormalTok{  \{ }\ControlFlowTok{return}\NormalTok{ p + i; \}}
\NormalTok{\};}
\end{Highlighting}
\end{Shaded}

\caption{An Accessor that provides an expression of non-aliasing semantics for mdspan.}
\label{restrict-accessor}
\end{figure}

\subsubsection{Accessor Use Case: Atomic Access}

Frequently in HPC applications, it is necessary to access a region of
memory atomically for only a small portion of its lifetime. Constructing
the entity to be atomic for the entire lifetime of the underlying
memory, as is done with \texttt{std::atomic}, may have unacceptable
overhead for many HPC use cases. As an entity that references a region
of memory for a subset of that memory's lifetime, \texttt{mdspan} is
ideally suited to be paired with a fancy reference type that expresses
atomic semantics (that is, all operations on the underlying memory are
to be performed atomically by the abstract machine). With the
introduction of \texttt{std::atomic\_ref} in C++20, all that is needed
is an accessor that customizes the reference type and provides an
\texttt{access} method that constructs such a reference. An
implementation of such an \texttt{Accessor} is shown in Figure
\ref{atomic-accessor}.

\begin{figure}[!h]

\begin{Shaded}
\begin{Highlighting}[]
\KeywordTok{template}\NormalTok{ \textless{}}\KeywordTok{class}\NormalTok{ T\textgreater{}}
\KeywordTok{struct}\NormalTok{ AtomicAccessor \{}
  \KeywordTok{using} \DataTypeTok{element\_type}\NormalTok{ = T;}
  \KeywordTok{using}\NormalTok{ pointer = T*;}
  \KeywordTok{using}\NormalTok{ reference = atomic\_ref\textless{}T\textgreater{};}
\NormalTok{  reference access(pointer p, }\DataTypeTok{ptrdiff\_t}\NormalTok{ i)}
    \AttributeTok{const} \KeywordTok{noexcept}
\NormalTok{  \{ }\ControlFlowTok{return}\NormalTok{ atomic\_ref\{ p[i] \}; \}}
\NormalTok{  pointer offset(pointer p, }\DataTypeTok{ptrdiff\_t}\NormalTok{ i)}
    \AttributeTok{const} \KeywordTok{noexcept}
\NormalTok{  \{ }\ControlFlowTok{return}\NormalTok{ p + i; \}}
\NormalTok{\};}
\end{Highlighting}
\end{Shaded}

\caption{An Accessor that provides an expression of atomic reference semantics for mdspan.}
\label{atomic-accessor}
\end{figure}

\subsubsection{Accessor Use Case: Bit-Packing}

Similar to the infamous
\texttt{std::vector\textless{}bool\textgreater{}}, the accessor
abstraction can be used to return a fancy reference type that references
individual bits packed into the bytes of underlying memory. (Unlike
\texttt{std::vector\textless{}bool\textgreater{}}, though,
\texttt{std::accessor\_basic\textless{}bool\textgreater{}} does not do
this by default.)

\subsubsection{Accessor Use Case: Strong Pointer Types for Heterogeneous
Memory Spaces}

Heterogeneity often requires a program to access multiple, potentially
disjoint memory spaces. Thus far, vendor-provided APIs for heterogeneity
have tended to represent this memory with plain-old raw pointers. An
important emerging paradigm in modern programming model design is
so-called ``strong types'' (also called ``opaque typedefs'' or ``phantom
types''),\cite{strongTypes,phantomOrigin} wherein meaning is opaquely
attached to the form of the type. For instance,
\texttt{distance\textless{}double\textgreater{}} and
\texttt{temperature\textless{}double\textgreater{}} might be different
concrete types that the compiler forbids mixing, even though they both
would use \texttt{double} for storage and arithmetic. Applied to
heterogeneity, the paradigm would suggest replacing raw pointers with
opaque typedefs indicating things like their compatibility or
accessibility. This not only introduces safety with respect to memory
access by an execution resource, but also allows generic software design
strategies where execution mechanisms can be deduced from the type of
the data structure. In \texttt{mdspan}, such strong typing can be
injected via the customization of the associated pointer type in the
\texttt{Accessor}. Initially, of course, such extensions will be outside
of the C++ Standard (e.g., OpenMP, HIP, SYCL, and older versions of
CUDA), but this design provides a means of forward compatibility if and
when the Standard addresses the concept of heterogeneous memory
resources natively in C++.

%================================================================================%

%================================================================================%
\section{Implementation}

This work is accompanied by a production-oriented implementation of the
proposed \texttt{std::mdspan}, available at
\href{https://github.com/kokkos/mdspan}{github.com/kokkos/mdspan}, the
details of which are discussed in this section. While the proposal it
implements targets C++23, the implementation includes compatibility
modes for C++17, C++14, and C++11. The implementation also includes a
couple of macros, \texttt{MDSPAN\_INLINE\_FUNCTION} and
\texttt{MDSPAN\_FORCE\_INLINE\_FUNCTION}, that can be used to add the
appropriate markings to functions and function templates, such as
\texttt{\_\_device\_\_} for CUDA compatibility. It has been tested on
various versions of numerous C++ compilers, including GCC, Clang,
Intel's ICC, Microsoft's MSVC, and NVIDIA's NVCC. The implementation
modestly extends the proposal in several places (mostly with respect to
typos in the latter), all of which are documented in the implementation
repository (link given above).

%================================================================================%

%================================================================================%
\section{Benchmarks}

A common complaint about C++ abstractions in HPC is that they hinder
compiler optimizations. While that was largely true in the past, a
number of developments have improved the situation. More recent C++
standards introduce capabilities and constraints which help the compiler
optimize code. Furthermore, with the widespread adoption of C++
abstraction layers in industry, significant work has gone into
optimizing commonly used compilers. To demonstrate that \texttt{mdspan}
does not introduce overheads compared to using raw pointers with manual
indexing, we will show benchmark results both for the version using
\texttt{mdspan} and an equivalent implementation using raw pointers.
Since the difference in most benchmarks is very small, most figures in
this section show overhead of the \texttt{mdspan} version over the raw
pointer variant. Negative overhead indicates cases where the
\texttt{mdspan} version was faster.

\begin{figure*}[!ht]
\centering
\includegraphics[width=0.95\textwidth]{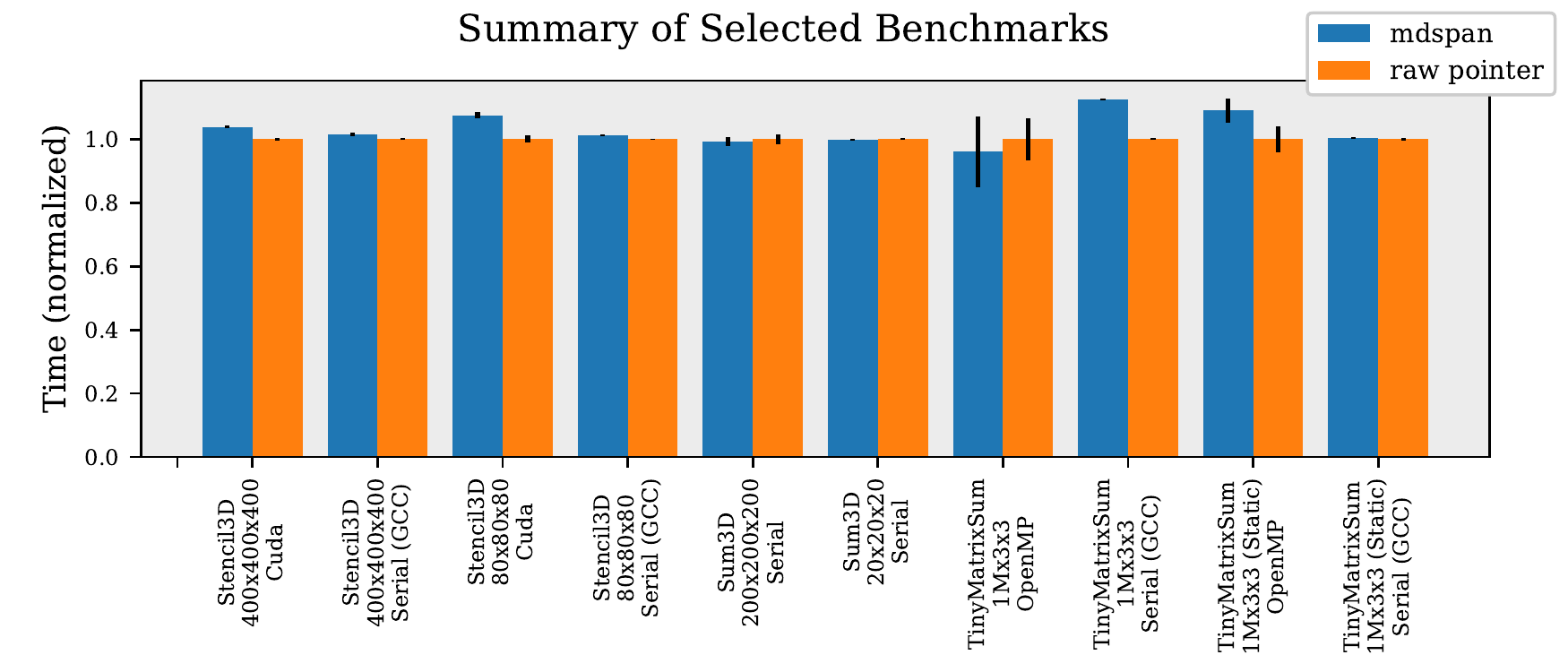}
\caption{An overview of selected benchmark comparisons of mdspan and raw pointer performance.  Each benchmark is normalized to the average execution time of the raw pointer case.  Details of each of these benchmarks are described in the text.}
\label{raw-vs-mdspan-overview}
\end{figure*}

Figure \ref{raw-vs-mdspan-overview} shows a normalized comparison of
\texttt{mdspan} versions of several selected benchmarks with the same
benchmark expressed with raw pointers. A more thorough elaboration
follows. Most of the benchmarks showed overheads within the measurement
noise, and no benchmarks showed overhead greater than 10\%. Examination
of generated assembly (and, at least in the case of the Intel compiler,
optimization reports) indicates similar---usually
identical---vectorization of the \texttt{mdspan} and raw pointer
versions of our benchmarks.

\subsection{Methodology}

All benchmarks were prepared and executed using the Google Benchmark
microbenchmarking support library.\cite{googlebenchmark} Table
\ref{machines} lists the test systems and compilers used for
benchmarking. Unless otherwise stated, CPU benchmarks were run on
Mutrino, and GPU benchmarks were run on Apollo. CPU benchmarks are
serial unless labeled ``OpenMP'', in which case they were parallelized
with the OpenMP \texttt{parallel\ for} directive on the outermost loop
(with the intent of measuring typical basic usage of OpenMP). CPU
benchmarks were compiled with GCC 8.2.0, Intel ICC 18.0.5, and the
latest Clang development branch (GitHub hash \texttt{1fcdcd0}, which is
LLVM SVN revision 370135; labeled as ``Clang 9 (develop)'' herein). GPU
benchmarks were compiled with NVIDIA's NVCC version 10.1, using GCC
5.3.0 as the host compiler. The source code of all benchmarks is
available on the mdspan implementation repository that accompanies this
paper (see Implementation section above). A brief description of each
benchmark is also included here for completeness. These benchmarks tend
to focus on the three-dimensional use case (which we view as the
smallest ``relatively non-trivial'' number of dimensions), but spot
checks with larger numbers of dimensions---up to 10---yielded similar
results and led to similar conclusions.

\begin{table}[htbp]
\caption{Test Systems and Software}
\begin{center}
%\rowcolors{1}{white}{black!10!white}
\renewcommand{\arraystretch}{1.5}
\begin{tabularx}{0.45\textwidth}{|c|>{\raggedright\arraybackslash}X|>{\raggedright\arraybackslash}X|}
\hline
Machine & Hardware & Compilers \\
\hline
    Apollo & NVIDIA TitanV & CUDA~10.1~(nvcc),\newline GCC~5.3.0 \\
Astra & Cavium Thunder-X2 CN9975-2000 & GCC~8.2~(ARM) \\ 
Blake & Intel(R) Xeon(R) Platinum 8160 & Intel~19.3.199, Intel~18.0.5 \\
Mutrino & Intel(R) Xeon(R) CPU E5-2698 v3 & Intel~18.0.5, GCC~8.2.0, Clang~9~(develop) \\
Waterman & IBM Power9 10C 80HT & GCC~7.3.0 \\
\hline
\end{tabularx}
\renewcommand{\arraystretch}{1.0}

\label{machines}
\end{center}
\end{table}

\subsubsection{\texorpdfstring{\texttt{Sum3D}
Benchmark}{Sum3D Benchmark}}

Intended as a ``simplest possible'' benchmark, this benchmark simply
sums over all of the entries in a 3D \texttt{mdspan}. The raw pointer
version (as with all of the benchmarks) does the same thing, but uses
hard-coded index arithmetic. Both right-most fast-running and left-most
fast-running loop structures and layouts were tested (and yielded
similar results), and only the right layout, right loop structure
results are discussed in this paper, for brevity. The relevant portion
of the source code for this benchmark, for an input \texttt{mdspan}
named \texttt{s} and an output named \texttt{sum}, looks like:

\begin{Shaded}
\begin{Highlighting}[]
\ControlFlowTok{for}\NormalTok{(}\DataTypeTok{ptrdiff\_t}\NormalTok{ i = }\DecValTok{0}\NormalTok{; i \textless{} s.extent(}\DecValTok{0}\NormalTok{); ++i) \{}
  \ControlFlowTok{for}\NormalTok{ (}\DataTypeTok{ptrdiff\_t}\NormalTok{ j = }\DecValTok{0}\NormalTok{; j \textless{} s.extent(}\DecValTok{1}\NormalTok{); ++j) \{}
    \ControlFlowTok{for}\NormalTok{ (}\DataTypeTok{ptrdiff\_t}\NormalTok{ k = }\DecValTok{0}\NormalTok{; k \textless{} s.extent(}\DecValTok{2}\NormalTok{); ++k) \{}
\NormalTok{      sum += s(i, j, k);}
\NormalTok{    \}}
\NormalTok{  \}}
\NormalTok{\}}
\end{Highlighting}
\end{Shaded}

\subsubsection{\texorpdfstring{\texttt{Stencil3D}
Benchmark}{Stencil3D Benchmark}}

This benchmark takes the sum of all of the neighboring points in
three-dimensional space from an input \texttt{mdspan} and stores it in
the corresponding entry of the output \texttt{mdspan}. In terms of
structured grid computations, it has a ``stencil size'' of one (which is
\texttt{d}, a \texttt{constexpr\ ptrdiff\_t} variable known to the
optimizer, in the code excerpt below). The relevant portion of the
source code for this benchmark, for an input \texttt{mdspan} named
\texttt{s} and an output \texttt{mdspan} named \texttt{o}, looks like
this:

\begin{Shaded}
\begin{Highlighting}[]
\ControlFlowTok{for}\NormalTok{(}\DataTypeTok{ptrdiff\_t}\NormalTok{ i = d; i \textless{} s.extent(}\DecValTok{0}\NormalTok{){-}d; ++i) \{}
  \ControlFlowTok{for}\NormalTok{(}\DataTypeTok{ptrdiff\_t}\NormalTok{ j = d; j \textless{} s.extent(}\DecValTok{1}\NormalTok{){-}d; ++j) \{}
    \ControlFlowTok{for}\NormalTok{(}\DataTypeTok{ptrdiff\_t}\NormalTok{ k = d; k \textless{} s.extent(}\DecValTok{2}\NormalTok{){-}d; ++k) \{}
      \DataTypeTok{value\_type}\NormalTok{ sum\_local = }\DecValTok{0}\NormalTok{;}
      \ControlFlowTok{for}\NormalTok{(}\DataTypeTok{ptrdiff\_t}\NormalTok{ di = i{-}d; di \textless{} i+d+}\DecValTok{1}\NormalTok{; ++di) \{}
        \ControlFlowTok{for}\NormalTok{(}\DataTypeTok{ptrdiff\_t}\NormalTok{ dj = j{-}d; dj \textless{} j+d+}\DecValTok{1}\NormalTok{; ++dj) \{}
          \ControlFlowTok{for}\NormalTok{(}\DataTypeTok{ptrdiff\_t}\NormalTok{ dk = k{-}d; dk \textless{} k+d+}\DecValTok{1}\NormalTok{; ++dk) \{}
\NormalTok{            sum\_local += s(di, dj, dk);}
\NormalTok{          \}}
\NormalTok{        \}}
\NormalTok{      \}}
\NormalTok{      o(i,j,k) = sum\_local;}
\NormalTok{    \}}
\NormalTok{  \}}
\NormalTok{\}}
\end{Highlighting}
\end{Shaded}

\subsubsection{\texorpdfstring{\texttt{TinyMatrixSum}
benchmark}{TinyMatrixSum benchmark}}

This benchmark applies a batched sum operation to large number of small
(in this paper, 3x3) matrices, accumulating from the input Nx3x3
\texttt{mdspan} into an Nx3x3 \texttt{mdspan}. The relevant portion of
the source code for this benchmark, for an input \texttt{mdspan} named
\texttt{s} and an output \texttt{mdspan} named \texttt{o}, looks like
this:

\begin{Shaded}
\begin{Highlighting}[]
\ControlFlowTok{for}\NormalTok{(}\DataTypeTok{ptrdiff\_t}\NormalTok{ i = }\DecValTok{0}\NormalTok{; i \textless{} s.extent(}\DecValTok{0}\NormalTok{); ++i) \{}
  \ControlFlowTok{for}\NormalTok{(}\DataTypeTok{ptrdiff\_t}\NormalTok{ j = }\DecValTok{0}\NormalTok{; j \textless{} s.extent(}\DecValTok{1}\NormalTok{); ++j) \{}
    \ControlFlowTok{for}\NormalTok{(}\DataTypeTok{ptrdiff\_t}\NormalTok{ k = }\DecValTok{0}\NormalTok{; k \textless{} s.extent(}\DecValTok{2}\NormalTok{); ++k) \{}
\NormalTok{      o(i,j,k) += s(i,j,k);}
\NormalTok{    \}}
\NormalTok{  \}}
\NormalTok{\}}
\end{Highlighting}
\end{Shaded}

\subsubsection{\texorpdfstring{\texttt{Subspan3D}
benchmark}{Subspan3D benchmark}}

This benchmark performs the same operations as the \texttt{Sum3D}
benchmark, but uses two calls to \texttt{subspan}, instead of accessing
the entries of the \texttt{mdspan} in the ``normal'' way
(\texttt{operator()} with three integer indices). It is intended to
stress the abstraction overhead (or lack thereof) in the implementation,
since \texttt{subspan} is the most complex part of the \texttt{mdspan}
implementation from a C++ perspective. Note that this is not the
intended use case of the \texttt{subspan} function, though it serves as
a reasonable worst-case proxy. The relevant portion of the source code
for this benchmark, for an input \texttt{mdspan} named \texttt{s} and an
output named \texttt{sum}, looks like this:

\begin{Shaded}
\begin{Highlighting}[]
\ControlFlowTok{for}\NormalTok{(}\DataTypeTok{ptrdiff\_t}\NormalTok{ i = }\DecValTok{0}\NormalTok{; i \textless{} s.extent(}\DecValTok{0}\NormalTok{); ++i) \{}
  \KeywordTok{auto}\NormalTok{ sub\_i = subspan(s, i, all, all);}
  \ControlFlowTok{for}\NormalTok{ (}\DataTypeTok{ptrdiff\_t}\NormalTok{ j = }\DecValTok{0}\NormalTok{; j \textless{} s.extent(}\DecValTok{1}\NormalTok{); ++j) \{}
    \KeywordTok{auto}\NormalTok{ sub\_i\_j = subspan(sub\_i, j, all);}
    \ControlFlowTok{for}\NormalTok{ (}\DataTypeTok{ptrdiff\_t}\NormalTok{ k = }\DecValTok{0}\NormalTok{; k \textless{} s.extent(}\DecValTok{2}\NormalTok{); ++k) \{}
\NormalTok{      sum += sub\_i\_j(k);}
\NormalTok{    \}}
\NormalTok{  \}}
\NormalTok{\}}
\end{Highlighting}
\end{Shaded}

\subsubsection{\texorpdfstring{\texttt{MatVec}
benchmark}{MatVec benchmark}}

The \texttt{MatVec} benchmark performs a simple dense matrix-vector
multiply operation. It is aimed at demonstrating the impact of layout
choice on performance, particularly in the context of performance
portability of parallelization across diverse hardware platforms.
Consider this serial implementation:

\begin{Shaded}
\begin{Highlighting}[]
\ControlFlowTok{for}\NormalTok{(}\DataTypeTok{ptrdiff\_t}\NormalTok{ i = }\DecValTok{0}\NormalTok{; i \textless{} A.extent(}\DecValTok{0}\NormalTok{); ++i) \{}
  \ControlFlowTok{for}\NormalTok{(}\DataTypeTok{ptrdiff\_t}\NormalTok{ j = }\DecValTok{0}\NormalTok{; j \textless{} A.extent(}\DecValTok{1}\NormalTok{); ++j) \{}
\NormalTok{    y(i) += A(i,j) * x(j);}
\NormalTok{  \}}
\NormalTok{\}}
\end{Highlighting}
\end{Shaded}

When parallelizing the outer loop via OpenMP, C++17 standard parallel
algorithms, or CUDA, the optimal layout depends on the hardware. On
CPUs, the compiler will vectorize the inner loop over \texttt{j}; thus,
unit-stride access on the second dimension of \texttt{A} is optimal. On
GPUs, no implicit auto-parallelization happens, so unit-stride access on
the first dimension is optimal. Being able to make this layout change in
the type of \texttt{A}---without actually changing the algorithm---means
that the algorithm can be generic over different architectures.

\subsection{Results: Compiler Comparison}

\begin{figure}[htbp]
\centering
\includegraphics[width=0.45\textwidth]{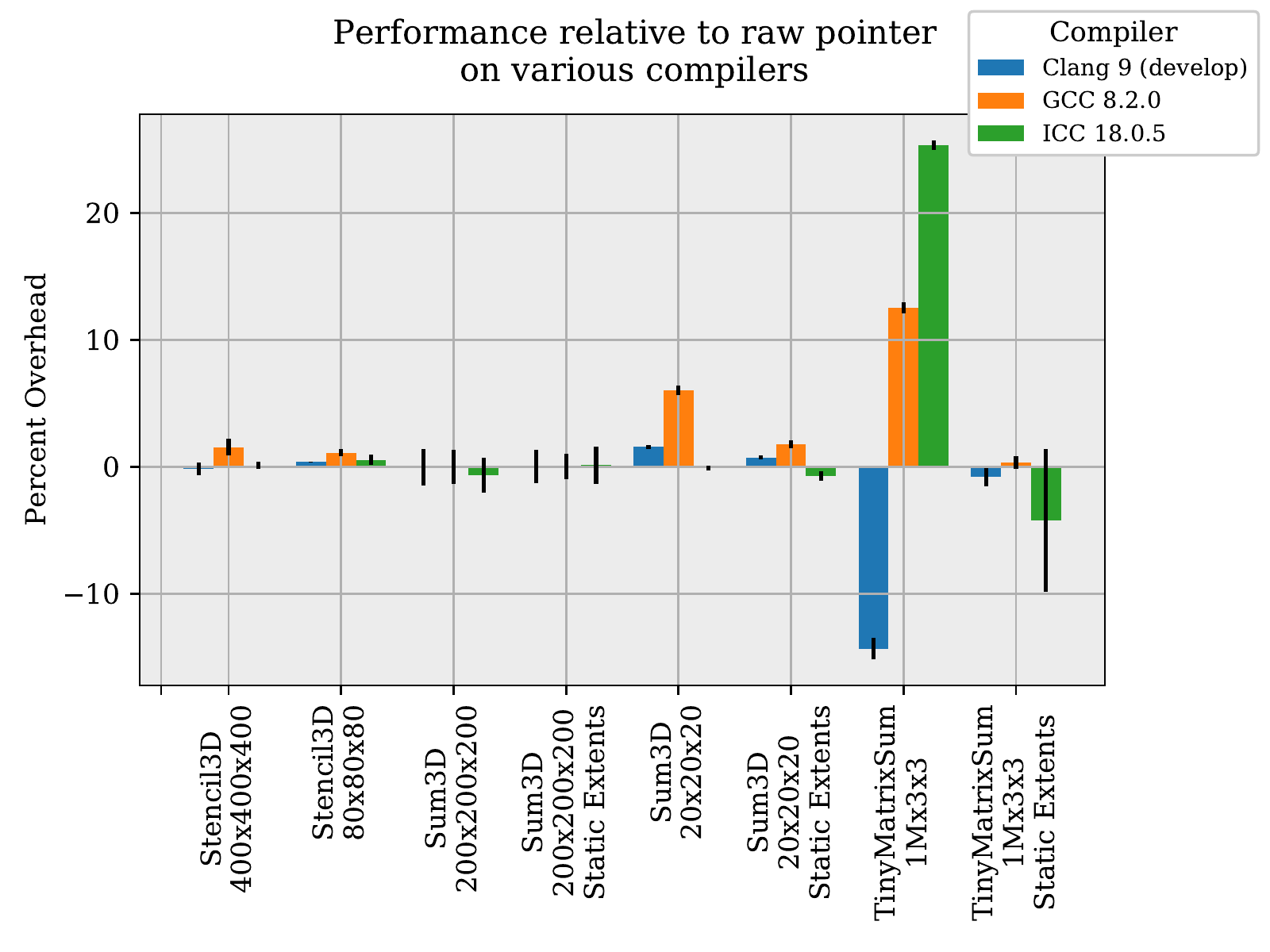}
\caption{Comparison of overheads, relative to raw pointer implementations, of the serial versions of various benchmarks across different compilers.}
\label{compiler-comparison}
\end{figure}

Figure \ref{compiler-comparison} shows a comparison of \texttt{mdspan}
overheads relative to the raw pointer analog for serial versions of
several benchmarks. With the exception of the \texttt{TinyMatrixSum}
benchmark using dynamic extents, overheads on all of the benchmarks were
either completely or very nearly within the experimental noise. The
outlier in this regard, \texttt{TinyMatrixSum} with dynamic extents, is
an interesting case study in the brittleness of modern loop optimizers,
whether or not C++ abstraction is involved. To a first approximation,
the authors believe the explanation for this is as follows: if the
compiler heuristic guesses that the inner loop sizes are too large, the
resulting optimization decisions (such as the amount of unrolling) are
inefficient for a 3x3 matrix. How the use of \texttt{mdspan} interacts
with the compiler's heuristic for generating this guess varies from
compiler to compiler. For instance, with the latest version of Clang,
the optimizer actually happens to make a \emph{better} guess, leading to
a ``negative overhead.''

In many ways, the optimizer brittleness in this single outlier presents
a strong argument for the sort of genericness that \texttt{mdspan}
provides. As C++ continues to evolve, more compiler-specific extensions
that let programmers give hints to guide compiler optimization are
likely to trickle in. Maintaining such hints inside the logic of
application code is often impractical or impossible, but incorporating
that information into the \texttt{mdspan} accessor (particularly if such
accessors can be vendor-provided), over which most algorithms can be
generic, is a completely reasonable proposition in many cases.

\subsection{Results: Effect of Static Extents}

\begin{figure}[htbp]
\centering
\includegraphics[width=0.45\textwidth]{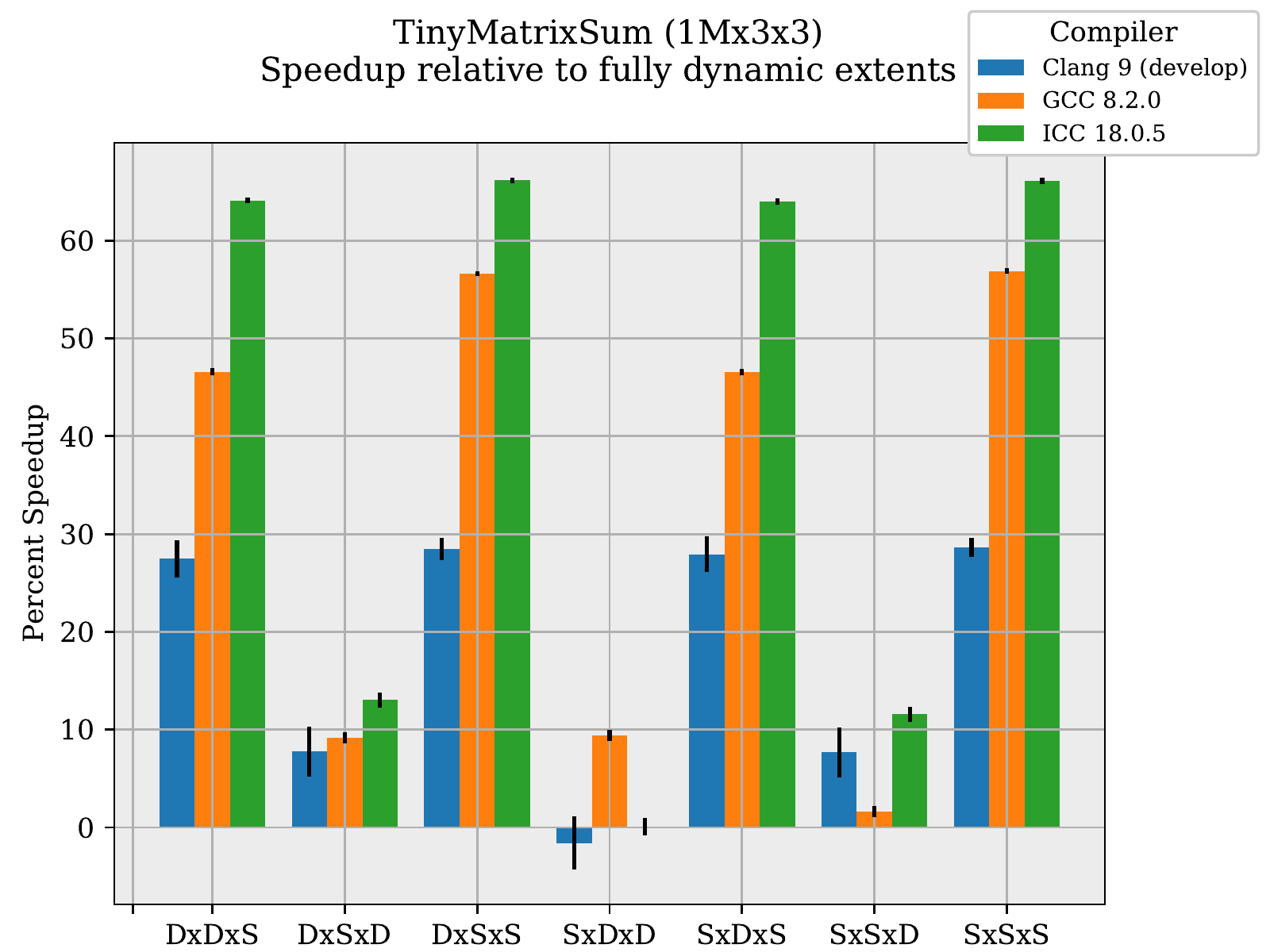}
\caption{Comparison of speedups, relative to the fully dynamic version, of the TinyMatrixSum benchmark.  "D" indicates a dynamic expression of the particular extent, while "S" indicates a static expression (for instance, "DxDxS" indicates that the first extent, 1 million, was expressed dynamically, the second extent, 3, was expressed dynamically, and the third extent, 3, was expressed statically).}
\label{static-extents}
\end{figure}

Figure \ref{static-extents} shows the speedup achieved when using static
extents for the two inner dimensions as opposed to dynamic extents. When
programmers provide them as static extents, the compiler is able to
unroll the inner loops fully, resulting in nearly two times better
performance on the test system Mutrino. The effect of static extents on
the compiler's ability to optimize can vary significantly from compiler
to compiler based on design decisions internal to the compiler's
implementation.

\subsection{Results: Effect of Layout Abstraction}

\begin{figure}[htbp]
\centering
\includegraphics[width=0.45\textwidth]{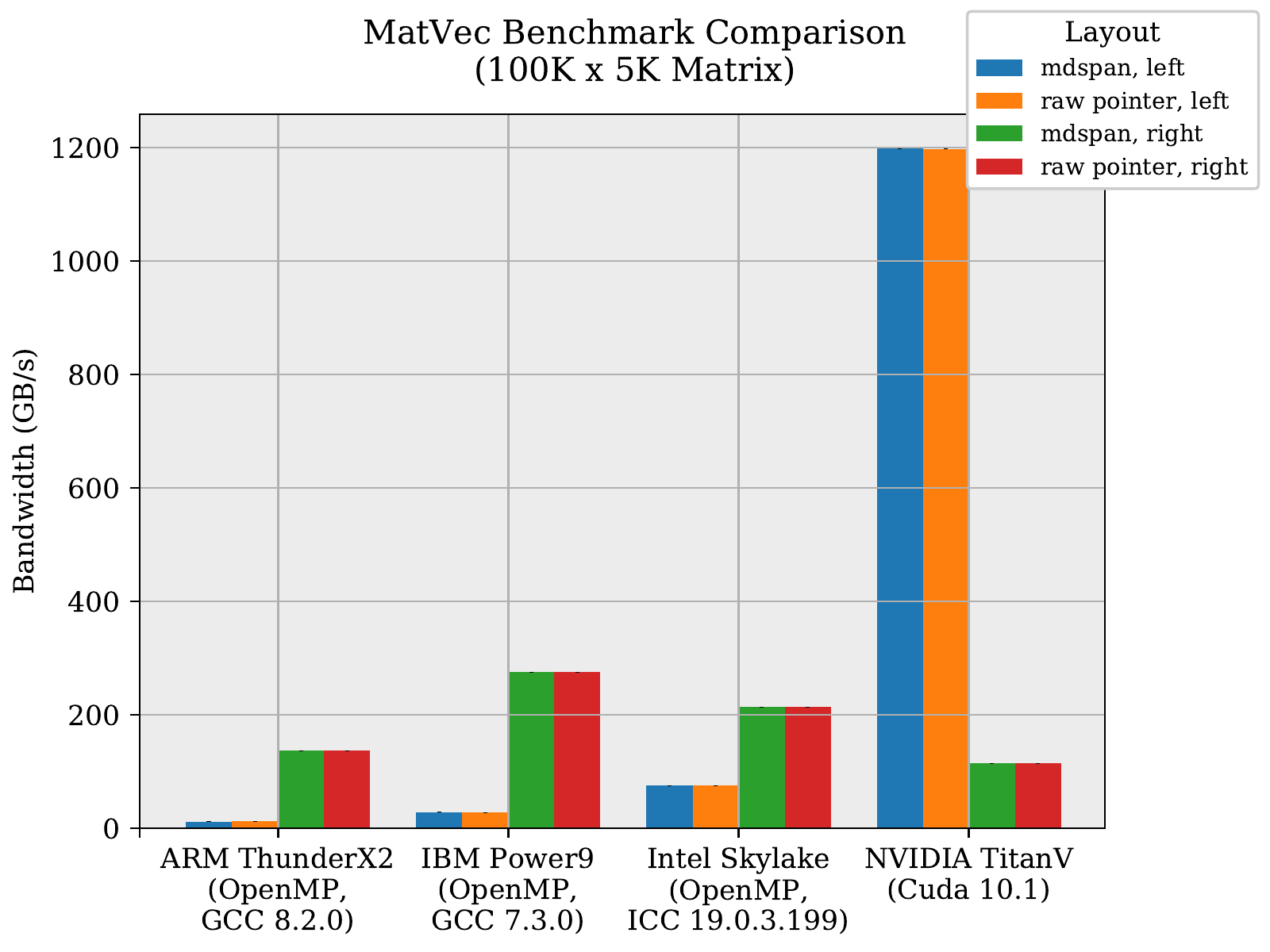}
\caption{Comparison of absolute memory bandwidths for the MatVec benchmark with different memory layouts.}
\label{layout-matvec}
\end{figure}

The benchmark in Figure \ref{layout-matvec} was run on the ARM ThunderX2
(test system Astra), Intel SkyLake (test system Blake), and NVIDIA
TitanV (test system Apollo) platforms using OpenMP parallelization for
the CPUs and CUDA for the GPU. On the CPU systems the use of
\texttt{layout\_right} (for the matrix) provides the better performance,
with \texttt{layout\_left} being 3x-7x slower. On the GPU, however, the
\texttt{layout\_left} version achieves a 10x higher throughput. The
results shown represent performance measured in terms of algorithmic
memory throughput (that is, the count of memory accesses in the
algorithm divided by run time.)

\subsection{\texorpdfstring{Results: Overhead of
\texttt{subspan}}{Results: Overhead of subspan}}

\begin{figure}[htbp]
\centering
\includegraphics[width=0.45\textwidth]{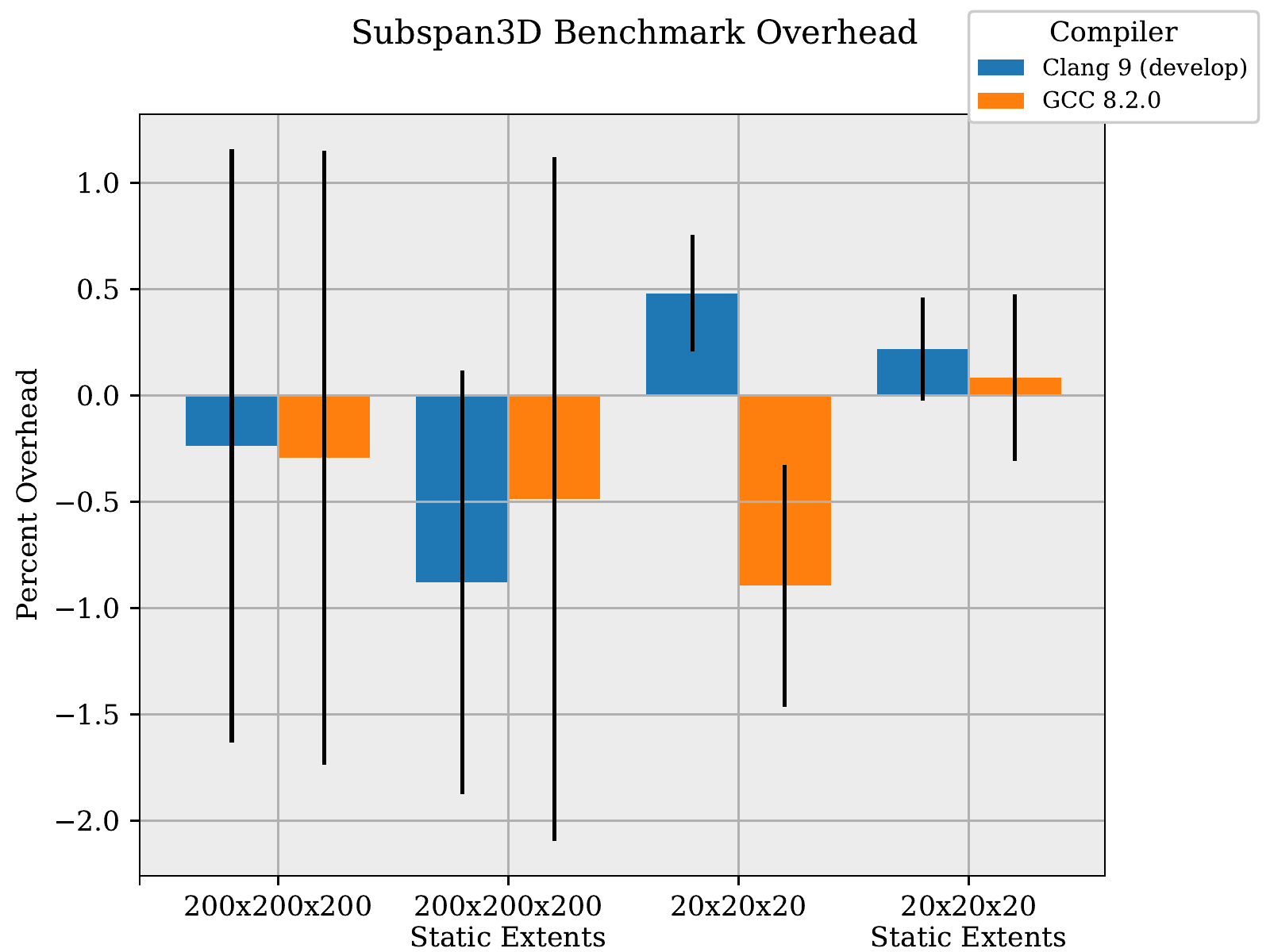}
\caption{Comparison of overheads, relative to raw pointer implementations, of the Subspan3D benchmark for GCC and Clang.}
\label{subspan-gcc-and-clang}
\end{figure}

\begin{figure}[htbp]
\centering
\includegraphics[width=0.5\textwidth]{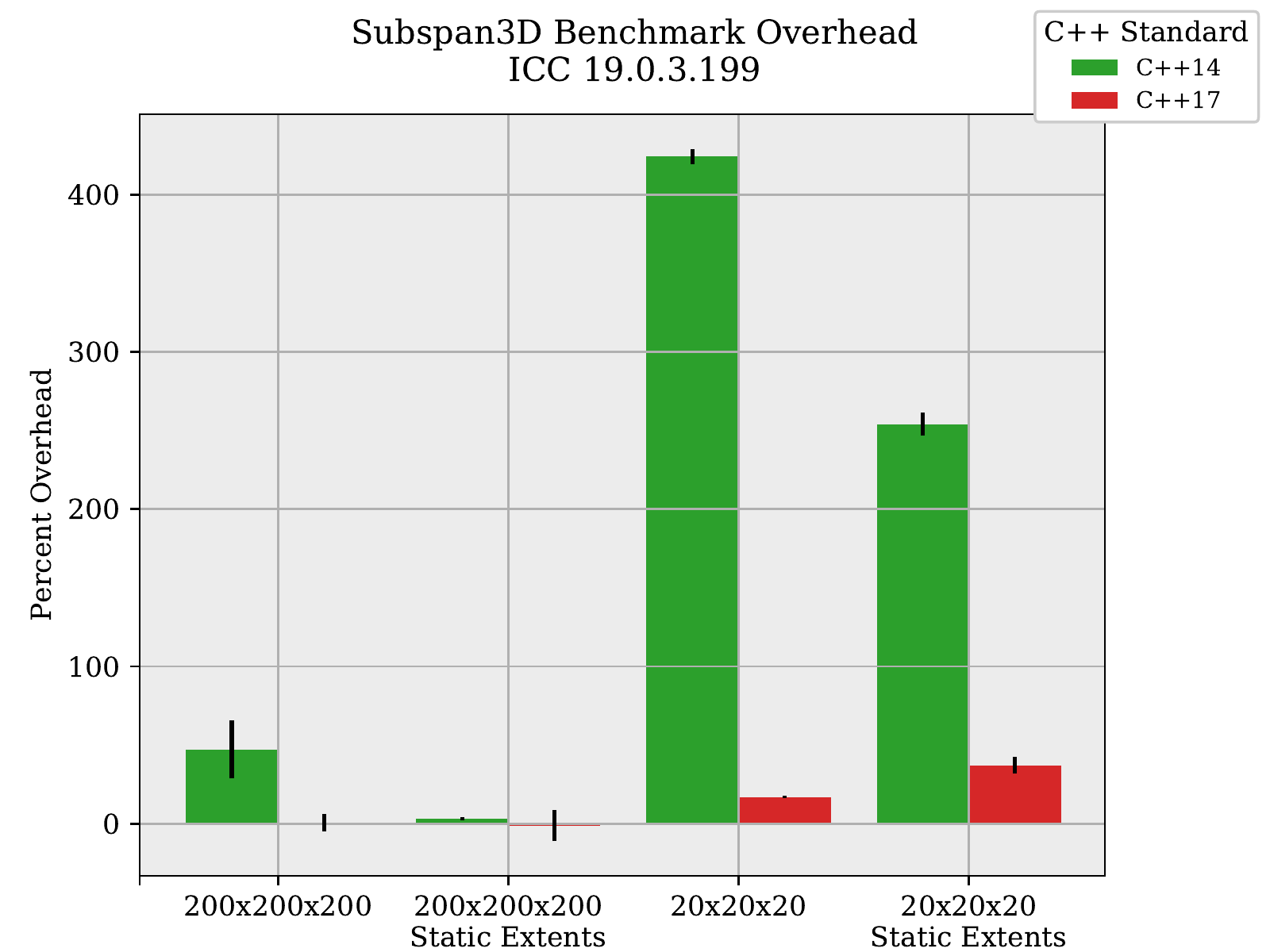}
\caption{Comparison of overheads, relative to raw pointer implementations, of the Subspan3D benchmark for ICC 19.0.3.199.  Note that this compiler was not available on our primary testing machine, so the test system Blake was used for this benchmark.}
\label{subspan-intel}
\end{figure}

For recent versions of GCC and Clang, the results are essentially
identical to the raw pointer implementation of \texttt{Sum3D}, as shown
in Figure \ref{subspan-gcc-and-clang}. (There is no raw pointer
implementation of \texttt{Subspan3D}, since the whole point is that it
would be identical to \texttt{Sum3D}.) For ICC 18.0.5, the results
showed significant overhead, rendering the GCC and Clang results
invisible---as much as 400\%. (The absolute magnitudes of the raw
pointer timings were similar across all three compilers, so this is a
genuine measurement of overhead introduced by the ICC frontend). Using
the more recent ICC 19.0.3.199, we were able to obtain much more
reasonable results in C++17 mode. Interestingly, though, the C++14
results \emph{with the same compiler version} were much more similar to
the ICC 18.0.5 results, indicating that the difference arises, at least
in part, from more modern C++ abstractions being easier for modern
compilers to understand. These results are shown in Figure
\ref{subspan-intel}.

%================================================================================%

%================================================================================%
\section{Conclusions}

Based on the \texttt{View} data structure in the Kokkos C++ Performance
Portable Programming Model, \texttt{mdspan} introduces a
multi-dimensional array view abstraction into the C++ Standard. We have
presented both the ISO C++ design and a production-oriented
implementation of \texttt{mdspan}. The design includes layout and
accessor abstractions that address performance portability concerns.
Besides controlling memory access patterns and data access semantics,
the abstractions also open the door for incorporating heterogeneous
memory paradigms via strong typing. Using a number of microbenchmarks,
we have demonstrated that our implementation of \texttt{mdspan} has (in
most cases) negligible overhead, compared to using raw pointers to
represent multi-dimensional arrays. The implementation can be used with
a C++11 standard-compliant compiler, and thus can be used with currently
available toolchains on typical supercomputing systems. The
standardization of \texttt{mdspan} lays the foundation for further
efforts, such as standardized linear algebra,\cite{wg21_p1673} which can
help to address future performance portability needs of HPC and
heterogeneous computing use cases.

\section{Acknowledgments}

This work was carried out in part at Sandia National Laboratories.
Sandia National Laboratories is a multimission laboratory managed and
operated by National Technology \& Engineering Solutions of Sandia, LLC,
a wholly owned subsidary of Honeywell International Inc., for the U. S.
Department of Energy's National Nuclear Security Administration under
contract DE-NA0003525.

\bibliographystyle{IEEEtran}
\bibliography{IEEEabrv,david,crtrott,bryce}
%\begin{thebibliography}{00}
%\bibitem{b1} G. Eason, B. Noble, and I. N. Sneddon, ``On certain integrals of Lipschitz-Hankel type involving products of Bessel functions,'' Phil. Trans. Roy. Soc. London, vol. A247, pp. 529--551, April 1955.
%\bibitem{b2} J. Clerk Maxwell, A Treatise on Electricity and Magnetism, 3rd ed., vol. 2. Oxford: Clarendon, 1892, pp.68--73.
%\bibitem{b3} I. S. Jacobs and C. P. Bean, ``Fine particles, thin films and exchange anisotropy,'' in Magnetism, vol. III, G. T. Rado and H. Suhl, Eds. New York: Academic, 1963, pp. 271--350.
%\bibitem{b4} K. Elissa, ``Title of paper if known,'' unpublished.
%\bibitem{b5} R. Nicole, ``Title of paper with only first word capitalized,'' J. Name Stand. Abbrev., in press.
%\bibitem{b6} Y. Yorozu, M. Hirano, K. Oka, and Y. Tagawa, ``Electron spectroscopy studies on magneto-optical media and plastic substrate interface,'' IEEE Transl. J. Magn. Japan, vol. 2, pp. 740--741, August 1987 [Digests 9th Annual Conf. Magnetics Japan, p. 301, 1982].
%\bibitem{b7} M. Young, The Technical Writer's Handbook. Mill Valley, CA: University Science, 1989.
%\end{thebibliography}
%\vspace{12pt}
%\color{red}
%IEEE conference templates contain guidance text for composing and formatting conference papers. Please ensure that all template text is removed from your conference paper prior to submission to the conference. Failure to remove the template text from your paper may result in your paper not being published.

\end{document}